\documentclass{sig-alternate}
\usepackage[applemac]{inputenc}
\usepackage{syntax}
\usepackage[cyr]{aeguill}
\usepackage{listings}
\usepackage{multirow}
\usepackage{enumitem}
\setlist[itemize]{leftmargin=*}
\usepackage{hyperref}

\lstdefinelanguage{SimpleDB} {morekeywords={MODEL,COMMIT,TABLE,PRIMARY,FOREIGN,KEY,REFERENCES,IF,THEN,ELSE,ENDIF,WHILE,DO,ENDWHILE,READ,LOAD,NEXT,CATCH,ROLLBACK,INSERT,INTO,VALUES,UPDATE,SET,WHERE,DELETE,FROM,TRUE,FALSE,NIL,SELECT,HEAD,TAIL,ENDMODEL},
sensitive=false}

\lstdefinelanguage{Alloy}
{morekeywords={abstract, all, and, as, assert, but, check, disj, else,
  exactly, extends, fact, for, fun, iden, if, iff, implies, in, Int,
  int, let, lone, module, no, none, not, one, open, or, part, pred,
  run, seq, set, sig, some, sum, then, univ},
sensitive=true,
morecomment=[l][\small\itshape]{--},
morecomment=[l][\small\itshape]{//},
morecomment=[s][\small\itshape]{/*}{*/},
tabsize=2,
columns=fullflexible,
literate={->}{{$\to$}}1 {^}{{$\mspace{-3mu}\widehat{\quad}\mspace{-3mu}$}}1
 {<}{$<$ }2 {>}{$>$ }2 {>=}{$\geq$ }2 {=<}{$\leq$ }2
 {<:}{{$<\mspace{-3mu}:$}}2 {:>}{{$:\mspace{-3mu}>$}}2
 {<=>}{{$\Leftrightarrow$ }}2 {=>}{{$\Rightarrow$ }}2 {+}{$+$ }2 {++}{{$+\mspace{-8mu}+$ }}2
 {\~}{{$\mspace{-3mu}\widetilde{\quad}\mspace{-3mu}$}}1
 {!=}{$\neq$ }2 {*}{${}^{\ast}$}1 {.}{$\cdot$}1
 {\#}{$\#$}1
}

\begin{document}
%
\conferenceinfo{Working Paper}{University of Namur}

\title{A Symbolic Execution Algorithm \\ for Constraint-Based Testing \\ of Database Programs }
%
%
%
%
%

\numberofauthors{3} 
%
\author{
%
%
\alignauthor
Michaël Marcozzi\titlenote{F.R.S.-FNRS Research Fellow}\\
       \affaddr{Faculty of Computer Science}\\
       \affaddr{University of Namur}\\
       \affaddr{Rue Grandgagnage, 21}\\
       \affaddr{Namur, Belgium}\\
       \email{mmr@info.fundp.ac.be}
\alignauthor
Wim Vanhoof\\
       \affaddr{Faculty of Computer Science}\\
       \affaddr{University of Namur}\\
       \affaddr{Rue Grandgagnage, 21}\\
       \affaddr{Namur, Belgium}\\
       \email{wva@info.fundp.ac.be}
\alignauthor
Jean-Luc Hainaut\\
       \affaddr{Faculty of Computer Science}\\
       \affaddr{University of Namur}\\
       \affaddr{Rue Grandgagnage, 21}\\
       \affaddr{Namur, Belgium}\\
       \email{jlh@info.fundp.ac.be}
}



\maketitle
\begin{abstract}
In so-called constraint-based testing, symbolic execution is a common technique used as a part of the process to generate test data for imperative programs. Databases are ubiquitous in software and testing of programs manipulating databases is thus essential to enhance the reliability of software. This work proposes and evaluates experimentally a symbolic execution algorithm for constraint-based testing of database programs. First, we describe SimpleDB, a formal language which offers a minimal and well-defined syntax and semantics, to model common interaction scenarios between programs and databases. Secondly, we detail the proposed algorithm for symbolic execution of SimpleDB models. This algorithm considers a SimpleDB program as a sequence of operations over a set of relational variables, modeling both the database tables and the program variables. By integrating this relational model of the program with classical static symbolic execution, the algorithm can generate a set of path constraints for any finite path to test in the control-flow graph of the program. Solutions of these constraints are test inputs for the program, including an initial content for the database. When the program is executed with respect to these inputs, it is guaranteed to follow the path with respect to which the constraints were generated. Finally, the algorithm is evaluated experimentally using representative SimpleDB models.
\end{abstract}




\section{Introduction}

Testing \cite{Jorgensen2008,testing-book-kaner} is a complex process which constitutes the primary approach to improve the reliability of software, motivating \cite{Ramler:2006} much research to develop relevant automated approaches to test the quality of all aspects of software. In this work, we consider the automation of test inputs generation for functional unit testing of database-driven software.

Many advanced techniques have been developed so far (see e.g. \cite{adequacy-zhu}) to automate the generation of adequate test data for proper testing of units of code, with regard to their expected functions, and independently of any interaction with a database. Many approaches have also been proposed (e.g. \cite{8,11,9,5,10}) to automate the generation of test database contents, to be used for testing databases and database management systems (DBMS), independently of the data-flow in the programs interacting with the database or DBMS under test. Databases are nowadays ubiquitous in software and many units of code interact intensively with a large, persistent and highly-structured relational database \cite{Date:2003}. But barely few works (see \cite{3,13,moi-1}) have studied how to automate the generation of test inputs for such database programs, to test the correct \emph{interaction} between the code and the database. \\

In this work, we propose a technique to generate simultaneously input database contents and program input values, in order to test an imperative program interacting with a relational database through SQL statements. The technique adopts a white-box and structural testing approach. Given a finite set of execution paths in the control flow graph of the program, satisfying a given code coverage criterion, it generates, for each path in the set, test inputs leading to the execution of this path, using static forward symbolic execution. 

Such symbolic execution was introduced by King \cite{King-1976}, Clarke \cite{Clarke-1976} and DeMillo \& Offut \cite{Offutt:1991} and has been advocated in many constraint-based test data generation techniques (e.g. \cite{Cadar06exe,francois-constraints,Gotlieb:1998,meudec2001atgen,Offutt:1999}), in the context of programs having no interaction with a database. Given a path through the program code to be tested, symbolic execution builds a set of path constraints over the program inputs. These constraints are such that when the program is executed with respect to input values satisfying them, the execution is guaranteed to follow the path to be tested. In order to build these constraints, symbolic execution considers the static single assignment form of the program and it expresses the control dependencies imposed by the execution of the particular execution path to be tested. 

The technique proposed in this work adapts symbolic execution to the particular case of database programs. The core strategy of the technique is, as we proposed in \cite{moi-1}, to model every variable of the program and every table (which is, fundamentally, a relation) in the database as a relational variable containing a mathematical relation over simple domains, like integers. Each statement in the program, including both imperative and SQL statements, can then be modeled as a simple operation over these relational variables. By applying the classical static symbolic execution mechanism over this relational version of the program, we can derive a set of path constraints over the program input variables. The generated constraints are here relational constraints and the input variables refer both to the classical inputs of the program and to the content of each of the database tables at the program start.

As some of the relational variables manipulated by the relational version of the program model tables in a database, they must obey the constraints described by the relational schema of this database, such as, for example, the primary key or foreign key constraints. The technique proposed in this work expresses these schema constraints as relational constraints as well, and the path and schema constraints can then be combined into an unique input constraints system.  Each solution to this relational constraints system represents a test input, including an initial state for each table in the database, with respect to which the program can be executed and is guaranteed to follow the execution path to be tested. \\

The two main contributions of this work can be summarized as follows. 

First, we propose a formal language, called SimpleDB, to facilitate the formal analysis of database programs. SimpleDB refines end extends the ImperDB language that we defined in \cite{moi-1}, introducing database schema specification within the language, transactions management, use of lists as input variables and a fully defined syntax and semantics. A SimpleDB \emph{model} describes both an imperative program to be tested and the schema of the relational database manipulated by this program. SimpleDB is a tiny formal language, offering only a minimal set of classical well-defined primitives necessary for building database programs. These simplicity and formalness allow to  specify algorithms able to automate testing of database programs in a simple, fast and rigorous way. But despite its simplicity, SimpleDB allows to model a large and interesting part of the possible interaction scenarios between a real program and a real database. Notably, SimpleDB proposes basic mechanisms for throwing and catching exceptions. This is an important aspect of the language, as it allows for a clean testing of all execution paths, including those that may lead to an erroneous interaction between the program and the database. 

Secondly, we introduce, detail and evaluate a complete algorithm based on symbolic execution to generate test inputs for SimpleDB models. This algorithm extends, concretizes and permits experimentation to validate the raw strategy that we suggested in \cite{moi-1}: given a SimpleDB model and a path in the program described by this model, the algorithm generates the corresponding relational input constraints system in the Alloy language \cite{alloy-book}. A prototype of the algorithm has been evaluated using sample SimpleDB models, and the generated constraints have been solved using the Alloy Analyzer \cite{alloy-book}, showing promising results.\\

The remainder of this paper is organized as follows.  Section \ref{simpleDB} details the syntax and semantics of the SimpleDB language. In section \ref{algorithm}, we describe and illustrate the symbolic execution algorithm, able to generate Alloy input constraints for any finite execution path in any SimpleDB model. We provide experimental results in section \ref{experimentation}, showing the efficiency of the algorithm over several sample SimpleDB models. Finally, some conclusions, related and future work are provided in section \ref{conclusion}.

\section{SimpleDB: a minimalist syntax \\ and semantics for DB programs}
\label{simpleDB}

In this section, we detail the syntax and semantics of the SimpleDB language using a step by step approach. For each step, a part of the BNF grammar, describing some of the syntactic constructs of the language, is presented. For each syntactic construct, additional syntactic rules are explained, as well as the construct's semantics. The major language design choices are discussed if needed. The chosen notation for the BNF grammar includes some additional meta-symbols: \{...\} (grouping), * (repetition zero or more times) and $^+$ (repetition one or more times). When a single nonterminal appears several times in a single production, subscript notation allows to distinguish between the occurrences. 

A sample SimpleDB model is provided in figure \ref{sample}. This model defines a database with two tables: one for authors and one for the theatrical plays these authors write. The number of plays written by an author is stored for each author. The model also defines a program manipulating this database: the program adds a set of new plays to the database and updates the authors' plays counts. If the author of an added play was not in the database, it is added to the database as well. The plays are inserted one by one in isolated transactions.

\begin{figure*}
\centering
\caption{SimpleDB model with program inserting plays and updating author's count in a plays database}
\lstset{language=SimpleDB,frame=single,columns=flexible}
{\small
\begin{lstlisting} 
MODEL example
TABLE author (name,numberOfPlays,PRIMARY KEY(name),numberOfPlays > 0);
TABLE play (title,theAuthor,PRIMARY KEY(title),FOREIGN KEY(theAuthor) REFERENCES author);
COMMIT();
LOAD(newPlays);
WHILE (!(newPlays=NIL)) DO
	error = 0;
	READ(authorName);
	authors = SELECT name FROM author WHERE (name = authorName);
	isEmpty = CATCH(NEXT(authors));
	IF (isEmpty=1) THEN
		INSERT INTO author VALUES (authorName,1);
	ELSE 
		UPDATE author SET numberOfPlays = (numberOfPlays + 1) WHERE (name = authorName);
	ENDIF;	
	error = CATCH(INSERT INTO play VALUES (newPlays.HEAD,authorName));
	IF (error=0) THEN
		COMMIT();
	ELSE
		ROLLBACK();
	ENDIF;
	newPlays = newPlays.TAIL;
ENDWHILE;		
COMMIT();
ENDMODEL
\end{lstlisting}}
\label{sample}
\end{figure*}

\subsection{Model}

\begin{grammar}
{\small
<model> ::= "MODEL" <id> <db-schema> <program> "ENDMODEL"

<id> ::= \{a | ... | z | A | ... | Z\}\{a | ... | z | A | ... | Z  | 0 | ... | 9\}*
}	
\end{grammar}	  

A SimpleDB model is given a name and details first the relational schema of the database and subsequently the code of the program working on this database.

\subsection{Database schema}

\begin{grammar}
{\small
<db-schema> ::= <table>*

<table> ::= "TABLE" <id> "(" <attrib>$^+$ <pr-key> <f-key>* <constr>* ")"";"
}	
\end{grammar}	  

The relational database schema is a list of table definitions. This list can be empty. In such a case, the program works independently of any database. Each table is identified by its name, contains at least one attribute and endorses exactly one primary key. Foreign keys and additional constraints can be declared for a table. Semantics of schema definition primitives in SimpleDB is the same as defined in the classical SQL DDL semantics \cite{sql}.

\begin{grammar}
{\small
<attrib> ::= <id>","

<pr-key> ::= "PRIMARY" "KEY" "(" <id> ")"

<f-key> ::= "," "FOREIGN" "KEY" "(" <id>$_{att}$ ")" "REFERENCES" <id>$_{tab}$

<constr> ::= "," <id> \{"<" | "=" | ">"\} <natural>

<natural> ::= \{1 | ... | 9\}\{0 | ... | 9\}$^+$ | \{0 | ... | 9\}
}	
\end{grammar}	  

Each attribute in a table is simply identified by its name and is of integer type. The exclusive use of integer values in SimpleDB models does not limit the expressive power of our model of database programs since all other usual primitive types such as booleans, strings, and floating point numbers, but also data structures such as sets, arrays and matrices, can easily be mapped to integers and/or simulated using lists of integers. It does, however, make both the modeling and the use of a constraint solver conceptually simpler. These last reasons explain why the examples of SimpleDB models used in this work manipulate values which are not usually integers, like author names in the example of table \ref{sample}. 

Among the attributes of a table, one is declared to be the primary key of the table. Any attribute can be be declared to be a foreign key referencing another table in the schema. An attribute can reference several different tables. A row in a table cannot be deleted or see its primary key value updated as long as there exists at least another row in the database that references it. Cycles in tables referencing are not allowed. Simple arithmetic constraints can be declared over each of the attributes of a table. 

\subsection{Program code}

SimpleDB allows to write imperative programs processing integers and lists of integers and interacting with a database through transactions involving one or several simple SQL Select, Insert, Update or Delete statements.

\begin{grammar}
{\small
<program> ::= "COMMIT""("")"";" <stmt>* "COMMIT""("")"";"
}	
\end{grammar}	 

The code of the program is a sequence of statements which starts and ends with a commit statement. Variables must not be declared in SimpleDB. A variable can be used in all the statements subsequent to its initialization through a read, load or assignment (\synt{id}\lit*{=} ...) statement. A variable cannot be used outside of the code block where it was initialized. This means that a variable initialized inside a loop body/if branch cannot be used in the statements outside of this loop body/if branch. A variable can be of three types: integer, list of integers or table. The type of a variable is fixed by its initialization statement and any type change in a subsequent statement will result in a compile-time error. SQL statements semantics in SimpleDB is the same as defined by the classical SQL DML semantics \cite{sql}. 

\subsubsection{Imperative statements and lists management} 

\begin{grammar}
{\small
<stmt> ::=  "IF" <cond> "THEN" <stmt>*$_{then}$ "ELSE" <stmt>*$_{else}$ "ENDIF" ";" 
		  \alt "WHILE" <cond> "DO" <stmt>* "ENDWHILE" ";" 
		 \alt <id> "=" <expr> ";" \hfill (Assignment of variable <id>)
		 
<cond> ::= "TRUE" \hfill (Logical true value)
		\alt "FALSE" \hfill (Logical false value)
		\alt "("<cond>$_1$  \{"&&" | "||"\} <cond>$_2$")" \hfill (Logical conjunction \\ $_{}$ \hfill and disjunction)
		\alt "(""!" <cond>")" \hfill (Logical negation)
		\alt "("<int-expr>$_1$ \{"<" | "=" | ">"\} <int-expr>$_2$")" \hfill (Arithmetic \\ $_{}$ \hfill comparison)
  		\alt "("<id> "=" "NIL"")" \hfill (List emptiness test for list in variable <id>)

<expr> ::= <int-expr> | <list-expr>
		
<int-expr> ::= <id> \hfill (Integer-typed variable)
		\alt <natural>  \hfill (Natural number)
		\alt  "("<int-expr>$_1$  \{"+" | "-" | "*" | "/"\} <int-expr>$_2$")"  \hfill (Arithmetics)
		\alt "(" "-" <int-expr> ")" \hfill (Unary minus)
		\alt <id>".""HEAD" \hfill (First element of list in variable <id>)

<list-expr> ::= <id> \hfill (List of integers-typed variable)
		\alt "NIL" \hfill (Empty list)
		\alt  "["<int-expr>  "," <list-expr>"]" \hfill (Appending integer <int-expr> \\ $_{}$ \hfill at the beginning of list <list-expr>)
		\alt <id>".""TAIL" \hfill (List in variable <id> without its first element)
}	
\end{grammar}	  

SimpleDB allows to control the program flow using classical condition and loop statements. Classical variable assignment statement evaluates an integer or list of integers expression and assigns the obtained value to a variable of the program. SimpleDB allows all basic logic operations in if and while conditions, as well as arithmetic comparisons and list emptiness testing. SimpleDB allows full arithmetics over integers and has the basic operations over lists of integers (concatenation of an element to a list and selecting the head, respectively, tail of a list). Lists are immutable.  

\subsubsection{Interacting with the outside world} 

\begin{grammar}
{\small		  
<stmt> ::=    "READ" "("<id>")" ";"  
		  \alt "LOAD" "("<id>")" ";"  
}	
\end{grammar}	  

Read (respectively Load) statement assigns an integer (respectively list of integers) value from the outside world to one of the variables of the program. This models different kinds of interaction between the program and the outside world (except from the interaction with the database) such as parameters received from a calling program, user prompt, network access, reading from a file, etc.

\subsubsection{Reading data from the database}

\begin{grammar}
{\small		  
<stmt> ::= <id>"=SELECT"\{<id>$_i$","\}*<id>$_n$ "FROM" <id>$_{tab}$ "WHERE" <db-cond>";"
		  \alt "NEXT" "("<id>")" ";"
		  \alt <id> "=" "CATCH" "(""NEXT" "("<id>")"")" ";"
		   
<db-cond> ::= "TRUE"
		\alt "FALSE"
		\alt "("<db-cond>$_1$  \{"&&" | "||"\} <db-cond>$_2$")"
		\alt "(""!" <db-cond>")"
		\alt "("<id> \{"<" | "=" | ">"\} <int-expr>")" \hfill (<id> refers to an attribute of \\ $_{}$ \hfill the table being read/modified)
}	
\end{grammar}	  

In order to access database data within the program, a SQL Select query must be processed over the database content and the table returned by this query must be assigned to a variable of the program. These two steps are executed by the \synt{id} = \lit*{SELECT} ... statement. Subsequently, the table in the assigned variable can be accessed by the program, but only one row at a time, using a cursor pointing at the single readable row of the table. After assigning a table to a variable, the Next statement must be called over this variable to set the cursor in front of the first row of the table in the variable. Every subsequent Next statement will move the cursor one row ahead. The integer value of attribute \synt{id}$_{att}$ in the row pointed to by the cursor of the table assigned to variable \synt{id}$_{tab}$ can be accessed using the following syntax:

\begin{grammar}
{\small	    
 <int-expr> ::=  <id>$_{tab}$"("<id>$_{att}$")"
}	
\end{grammar}	

If the cursor is in front of the last row of the table or if the Select query returned an empty table, every call to the Next statement will result in an exception to be thrown within the program.  A Next statement can be wrapped in a Catch statement. The later will set an integer-typed variable to 1 if an exception has been thrown by the statement it wraps, and to 0 otherwise. If an exception remains uncaught, the program immediately terminates, revealing a potential fault in the code or database design.

Unlike the classical SQL semantics and in order to avoid any non-deterministic behavior, the SimpleDB semantics requires that the tables returned by Select queries are always sorted by ascending order of the primary key attribute values. Put simply, the following SimpleDB query over the database in figure \ref{sample}:
\lstset{language=SQL,frame=single,columns=flexible,basicstyle=\small}
\begin{lstlisting} 
SELECT name,numberOfPlays 
FROM author 
WHERE (numberOfPlays > 10)
\end{lstlisting}
is always equivalent to the following SQL query :
\lstset{language=SQL,frame=single,columns=flexible,basicstyle=\small}
\begin{lstlisting} 
SELECT name,numberOfPlays 
FROM author 
WHERE (numberOfPlays > 10)
ORDER BY name
\end{lstlisting}

In the Where clause of a SQL statement, as well as in the Set clause of an Update SQL statement, \synt{id} in \synt{int-expr} can represent both a program variable and an attribute of the read or written table. In case of potential ambiguity, \synt{id} always represents the attribute in the table.  

\subsubsection{Writing data into the database}

\begin{grammar}
{\small
<stmt> ::= 	 <db-write> ";"
		  \alt <id> "=" "CATCH" "("<db-write>")" ";"

<db-write> ::="INSERT" "INTO" <id> "VALUES" "("\{<int-expr>$_i$","\}*<int-expr>$_n$")" 
		     \alt "UPDATE" <id>$_{tab}$ "SET" <id>$_{att}$"="<int-expr> "WHERE" <db-cond>
		     \alt "DELETE" "FROM" <id> "WHERE" <db-cond>
}	
\end{grammar}	  

SQL Insert, Update and Delete statements allow the program to write data into the database.  If the execution of a such a statement provokes a violation of one of the database schema constraints, the database remains unmodified by the statement and an exception is thrown within the program. As with Next statements, exceptions either can be  caught using a Catch statement or make the program terminate, revealing a potential fault in the code or database design.

\subsubsection{Transactions management}

\begin{grammar}
{\small	    
<stmt> ::= 	  "COMMIT""("")" ";"
		  \alt "ROLLBACK""("")" ";"  
}	
\end{grammar}	  

SQL transactions are managed through the classical Commit and Rollback statements. A Commit statement makes permanent all the changes made to the database by the program during the current transaction, closes this transaction and opens a new one. A Rollback statement restores the database to its state at the start of the current transaction, closes this transaction and opens a new one. In any SimpleDB program, a new transaction is started at program start, and all uncommitted changes are saved at program end, using a Commit statement. \newpage

\section{An algorithm for symbolic execution of SimpleDB programs}
\label{algorithm}

\subsection{Inputs and outputs}

The symbolic execution algorithm proposed here receives as inputs a SimpleDB model and an execution path in the program defined within this model. It produces as output a relational constraints system, whose solutions are such that when the program is executed with respect to any of these solutions, its execution will follow the given path. 

The execution path received as input by our algorithm is supposed to be a path in the program's control flow graph. In particular, it defines which branch of each of the encountered If statements was taken, how many times the body of each of the encountered loops was executed and which of the encountered Next and \synt{db-write} statements threw an exception. In the case of \synt{db-write} statements throwing exceptions, the path also specifies which database schema constraint caused the exception to be thrown. 

The constraints system generated as output by our algorithm allows to find values for the inputs of the input SimpleDB program. Inputs of a SimpleDB program are composed of the content of each database table defined by the model at program start, as well as all the values gathered from the outside world by the Read and Load statements executed by the program.
 
\subsection{Algorithm principle}

The principle of the algorithm is to perform a relational symbolic execution of the program path received as input. Each of the different values taken by the program variables and by the database tables during the execution of the path is represented by a corresponding symbolic relational variable. First, symbolic execution generates constraints stating that the variables corresponding to the initial content of each table in the database conform to the database schema. Then, symbolic execution analyzes one by one the program statements executed by the path, in the order in which the path specifies they are executed. Every time a statement sets or changes the value of a program variable or of a database table, symbolic execution generates a new constraint stating how the symbolic variable representing this new value can be computed as a function of the other symbolic variables. Every time a statement offers a choice in the way it can be executed that depends on the values of the program variables and of the database tables (conditions, loops, next and \synt{db-write} statements), symbolic execution generates a constraint over the symbolic variables such that when the program is executed with respect to values satisfying the constraint, the execution is guaranteed to take the path under consideration.  

Once all the statements encountered along the path have been analyzed, the set of relational constraints generated during analysis can be solved to find values for the relational symbolic variables that satisfy these constraints. If such a solution exists, the values of the symbolic variables corresponding to the program inputs constitute some test input data that will guarantee the execution of the considered execution path. If the constraints have no solution, then the considered path is infeasible. 

The generated relational constraints are expressed using a widely used and well-documented language, offering good analysis tools, called Alloy \cite{alloy-book}. 

\subsection{Symbolic variables and relational \\ constraints generation rules}

In this section, we illustrate the execution of the algorithm over the sample SimpleDB model detailed in figure \ref{sample}. We detail each step of the symbolic execution process over the path where the While loop is executed once, the Next statement throws an exception, the THEN branch of both the If statements are taken, and both the two Insert statements do not violate any database integrity constraint. At each step, we present the rules used by our algorithm to generate Alloy symbolic variables and constraints. This step by step rules description process allows us to introduce the whole symbolic execution mechanism of the algorithm. \\

The algorithm always starts by generating symbolic variables and relational constraints for the database tables defined within the model. For each table, an Alloy type is first defined so that every symbolic variable representing the content of the table will be of this type. Then a symbolic variable is created to represent the initial content of the table (1). Finally, constraints are generated to enforce on this content the primary key (2) as well as all the foreign keys (3) and arithmetic constraints (4) defined in the table. For the model of table \ref{sample}, the generated Alloy code is as follows. The reader should note that symbolic variables will be always created using the \textbf{sig} keyword followed by the name of the variable and by its type. Here a symbolic variable \emph{authorINPUTDB2} was defined to represent the initial content of the table AUTHOR, and a symbolic variable \emph{playINPUTDB1} was defined to represent the initial content of the table PLAY.  Relational constraints will be always generated using the \textbf{fact} keyword.

\begin{lstlisting}[language=Alloy,columns=flexible,escapechar=\%]
module example // Name of the Alloy constraints model
// START of Alloy type definition for table AUTHOR
sig author{name : Int,numberOfPlays : Int}
pred equalauthor[a:author,b: author] 
{a.name = b.name && a.numberOfPlays = b.numberOfPlays}
fact{all disj a,b: author | !equalauthor[a,b]}
// END of Alloy type definition for table AUTHOR

(1) sig authorINPUTDB2 in author {} 
(2) fact{all disj a,b:authorINPUTDB2 | !((a.name=b.name))} 
(4) fact{all a: author | a.numberOfPlays > 0} 

// START of Alloy type definition for table PLAY
sig play{theAuthor : Int,title : Int}
pred equalplay[a:play,b: play]
{a.theAuthor = b.theAuthor && a.title = b.title}
fact{all disj a,b: play | !equalplay[a,b]}
// END of Alloy type definition for table PLAY

(1) sig playINPUTDB1 in play {}
(2) fact{all disj a,b: playINPUTDB1 | !((a.title = b.title))} 
(3) fact{all a: playINPUTDB1 | 
			one b:authorINPUTDB2 |a.theAuthor = b.name} 
\end{lstlisting}

The second step executed by our algorithm is to generate a relational Alloy type for lists, which are not supported by default in Alloy:

\begin{lstlisting}[language=Alloy,columns=flexible,escapechar=\%]
one sig Nil {}
sig List {head: Int,tail: List + Nil}
\end{lstlisting}

The algorithm can then proceed with symbolic execution of the program defined in the model. The algorithm considers each statement one by one and follows the path received as input. In the case of the example model and path considered in this section, the Load statement is symbolically executed first. Symbolic execution for Load simply creates a new symbolic variable of type list to represent the loaded value:

\begin{lstlisting}[language=Alloy,columns=flexible,escapechar=\%]
one sig newplaysINPUTPROG1 in List + Nil {}
\end{lstlisting}

The second statement in the path is a While statement. As the path specifies that the loop body must be executed this time, a relational constraint is generated to specify that the loop condition should be true. In this case, it means that the current content (represented by the symbolic variable \emph{newplaysINPUTPROG1}) of the SimpleDB variable \emph{newPlays} should not be the empty list:

\begin{lstlisting}[language=Alloy,columns=flexible,escapechar=\%]
fact{!(newplaysINPUTPROG1 = Nil)}
\end{lstlisting}

Then the algorithm proceeds with symbolic execution of the statements in the loop body, as specified within the input path. The first statement is an Assignment statement. Symbolic execution for Assignment creates a new symbolic variable of the correct type to represent the new value of the assigned variable and generates a constraint to specify that this new symbolic variable contains the value that can be computed by evaluating the expression on the right of the "=" symbol:

\begin{lstlisting}[language=Alloy,columns=flexible,escapechar=\%]
one sig errorINTERNALPROG1 in Int {}
fact{errorINTERNALPROG1 = 0}
\end{lstlisting}

Symbolic execution for Read simply creates a new symbolic variable of type integer to represent the read value:

\begin{lstlisting}[language=Alloy,columns=flexible,escapechar=\%]
one sig authornameINPUTPROG2 in Int {}
\end{lstlisting}

Symbolic execution for Select creates a new symbolic variable of the type of the table on which the Select query is executed, to represent the Select result table. A relational constraint is then generated to specify that a row is part of the Select result table if and only if it is part of the current content of the table on which the Select query is executed and that it enforce the WHERE condition of the Select query:

\begin{lstlisting}[language=Alloy,columns=flexible,escapechar=\%]
sig authorsINTERNALPROG2 in author {}
fact{all e:author | (e in authorINPUTDB2 
		&& (e.name = authornameINPUTPROG2)) 
		<=> e in authorsINTERNALPROG2}
\end{lstlisting}

Symbolic execution for Catch first proceeds with symbolic execution of the statement wrapped by the Catch and then acts as an Assignment (2) statement. Symbolic execution for Next will depend on the cursor state (which must be stored by the algorithm) of the "nexted" SimbleDB variable and on whether the Next statement should throw an exception or not according to the input path:
\begin{enumerate}
\item No call to Next made before: 
\begin{enumerate}
\item \underline{No exception is thrown}: Add a relational constraint stating that the symbolic variable corresponding to the current content of the "nexted" SimbleDB variable should contain at least one element.
\item \underline{An exception is thrown}: Add a relational constraint stating that the symbolic variable corresponding to the current content of the "nexted" SimbleDB variable should contain no element. (1)
\end{enumerate}
\item Call(s) to Next made before and no exception thrown:
\begin{enumerate}
\item \underline{No exception is thrown}: Create a new symbolic variable to represent the content of the "nexted" SimpleDB variable. Generate a relational constraint stating that this new variable can be obtained from the old one by removing the element with the lowest primary key value from the old variable. Generate a relational constraint stating that the newly created symbolic variable should contain at least one element. 
\item \underline{An exception is thrown}: Add a relational constraint stating that the symbolic variable corresponding to the current content of the "nexted" SimbleDB variable should contain exactly one element.
\end{enumerate}
\item Last call to Next threw an exception:
\begin{enumerate}
\item \underline{No exception is thrown}: This is not allowed by the SimpleDB semantics.
\item \underline{An exception is thrown}: Do nothing.
\end{enumerate}
\end{enumerate}

In the example considered here, the algorithm chose option 1.b:

\begin{lstlisting}[language=Alloy,columns=flexible,escapechar=\%]
(1) fact{#authorsINTERNALPROG2=0}
(2) one sig isemptyINTERNALPROG3 in Int {}
(2) fact{isemptyINTERNALPROG3=1}
\end{lstlisting}

As the path specifies that the THEN branch of the IF statement must be executed this time, a relational constraint is generated to specify that the If condition should be true. In this case, in means that the current value (represented by the symbolic variable \emph{isemptyINTERNALPROG3}) of the SimpleDB variable \emph{isEmpty} should be one:

\begin{lstlisting}[language=Alloy,columns=flexible,escapechar=\%]
fact{(isemptyINTERNALPROG3 = 1)}
\end{lstlisting}

Symbolic execution for Insert creates a new symbolic variable (1) for the new content of the table on which the Insert statement is executed. Then a relational constraint is generated stating that this new variable can be obtained by adding one row with the correct attribute values to the old one (2). Finally, relational constraints are added to specify that no constraint was violated during insert. In the example considered here, a relational constraint is added to state that there should not be any row in the previous content of the table whose primary key value is the same as in the row inserted by the statement (3):

\begin{lstlisting}[language=Alloy,columns=flexible,escapechar=\%]
(1) sig authorINTERNALDB1 in author {}
(2) fact{one e:author | 
		authorINTERNALDB1=authorINPUTDB2+e 
		&& e.numberOfPlays = 1 
		&& e.name = authornameINPUTPROG2}
(3) fact{no e:authorINPUTDB2 | 
		e.name=authornameINPUTPROG2}
\end{lstlisting}

Here again, symbolic execution for Next statement first proceeds with symbolic execution of the wrapped statement and then acts as an Assignment statement (5). Symbolic execution for this Insert statement acts identically as for the previous Insert (1)(2)(3), but a relational constraint is added as well, stating that the newly inserted row references an existing row in the current content of the Author table (4). 

Table \ref{dbwrite} details the symbolic variables and relational constraints generated by the algorithm for every possible behavior of an Insert, Update or Delete statement. Table \ref{alloyOf} describes the rules used by the algorithm to translate between SimpleDB and Alloy expressions and conditions. Table \ref{abbrev} explains the abbreviations defined in table \ref{dbwrite} and \ref{alloyOf}.

\begin{lstlisting}[language=Alloy,columns=flexible,escapechar=\%]
(1) sig playINTERNALDB2 in play {}
(2) fact{one e:play | playINTERNALDB2=playINPUTDB1+e 
		&& e.theAuthor = authornameINPUTPROG2 
		&& e.title = newplaysINPUTPROG1.head}
(3) fact{no e:playINPUTDB1 | 
		e.title=newplaysINPUTPROG1.head}
(4) fact{one e:authorINTERNALDB1 | 
		e.name=authornameINPUTPROG2}
(5) one sig errorINTERNALPROG4 in Int {}
(5) fact{errorINTERNALPROG4=0}
\end{lstlisting}

As the path specifies that the THEN branch of the IF statement must be executed this time, a relational constraint is generated to specify that the If condition should be true:

\begin{lstlisting}[language=Alloy,columns=flexible,escapechar=\%]
fact{(errorINTERNALPROG4 = 0)}
\end{lstlisting}

Symbolic execution for Commit statements simply does nothing. Symbolic execution for Rollback statements tells the algorithm to use the symbolic variable representing the content of each database table before the last executed Commit statement (saved by the algorithm) to represent the current content of the table.

Symbolic execution for Assignment creates a new symbolic variable of the correct type to represent the new value of the assigned variable and generates a constraint to specify that this variable contains the value that can be computed by evaluating the expression on the right of the "=" symbol:

\begin{lstlisting}[language=Alloy,columns=flexible,escapechar=\%]
one sig newplaysINTERNALPROG5 in List + Nil {}
fact{newplaysINTERNALPROG5=newplaysINPUTPROG1.tail}
\end{lstlisting}

As the path specifies that the loop body must not be executed any more, a relational constraint is generated to specify that the loop condition should be false. 

\begin{lstlisting}[language=Alloy,columns=flexible,escapechar=\%]
fact{!(!(newplaysINTERNALPROG5 = Nil))}
\end{lstlisting}

As all the statements in the path have been symbolically executed, the algorithm stops and returns the generated Alloy constraints model.  The Alloy analyzer \cite{alloy-book} can be asked to find a solution for these constraints using the following commands:

\begin{lstlisting}[language=Alloy,columns=flexible,escapechar=\%]
assert inputsExist {!(newplaysINPUTPROG1 in List + Nil 
		&& authornameINPUTPROG2 in Int 
		&& authorINPUTDB2 in author 
		&& playINPUTDB1 in play) }
check inputsExist
\end{lstlisting}

The solution returned by the Alloy analyzer for the symbolic variables representing the input values of the program constitute input data (in this case, two empty initial contents for the author and play tables, one list containing only the integer 7 as input for the LOAD statement and the integer 7 as input for the READ statement) for the program which guarantee that the considered path will be executed: \\

\noindent\{authorINPUTDB2=\{\}, playINPUTDB1=\{\}, newplaysINPUTPROG1=[7,NIL], authornameINPUTPROG2=7\}

\lstset{language=Alloy,frame=none,escapechar=\%}
\begin{table*}
\centering
\caption{Symbolic variables and relational constraints generation for \synt{db-write} statements}
\begin{tabular}{|p{0.20\textwidth}|p{0.75\textwidth}|} \hline
\synt{db-write} ::= 

\lit*{INSERT INTO} \synt{id}

\lit*{VALUES(}\synt{int-expr}$_1$,...,
\synt{int-expr}$_i$,...,\synt{int-expr}$_n$\lit*{)} 
& \vspace{-4mm} \begin{lstlisting}[language=java,columns=flexible,escapechar=\%]
if (%no exception thrown in path for this INSERT%) { 
\end{lstlisting}  \vspace{-5mm} \begin{lstlisting}[language=Alloy,columns=flexible,escapechar=\%]
%\quad\quad%sig %\textsf{freshAlloyVar}% in %\synt{id}% {}
%\quad\quad%fact { one e:%\synt{id}% | %\textsf{freshAlloyVar}%=%\textsf{alloyName}+e% && e.%\textsf{att}%%$_*$% = alloyOf(%\synt{int-expr}$_*$%) }
%\quad\quad%fact { no e:%\textsf{alloyName}% | e.%\textsf{pk}% =alloyOf(%\synt{int-expr}$_{\textsf{pk}^{pos}}$%) } // Primary key is not violated
%\quad\quad%fact { one e:%\textsf{fk}%%$_*^{tab}$% | e.%\textsf{fk}%%$_*^{pk}$% = alloyOf(%\synt{int-expr}$_{\textsf{fk}_*^{pos}}$%)  } // Every foreign key is not violated
\end{lstlisting}\vspace{-4mm} \begin{lstlisting}[language=java,columns=flexible,escapechar=\%]
} else { if (%Exception thrown in path: inserted primary key value already exists in the table%) { 
\end{lstlisting}  \vspace{-4mm} \begin{lstlisting}[language=Alloy,columns=flexible,escapechar=\%]
%\quad\quad\quad\quad\quad%fact { one e:%\textsf{alloyName}% | e.%\textsf{pk}% =alloyOf(%\synt{int-expr}$_{\textsf{pk}^{pos}}$%) }
\end{lstlisting}\vspace{-4mm} \begin{lstlisting}[language=java,columns=flexible,escapechar=\%]
%\quad%%\quad%%\quad%} else if (%Exception thrown in path: an inserted foreign key value references no row%) {
\end{lstlisting}  \vspace{-4mm} \begin{lstlisting}[language=Alloy,columns=flexible,escapechar=\%]
%\quad\quad\quad\quad\quad%fact { no e:%\textsf{fk}%%$_i^{tab}$% | e.%\textsf{fk}%%$_i^{pk}$% = alloyOf(%\synt{int-expr}$_{\textsf{fk}_i^{pos}}$%)  }
\end{lstlisting}\vspace{-4mm} \begin{lstlisting}[language=java,columns=flexible,escapechar=\%]
%\quad%%\quad%%\quad%} else if (%Exception thrown in path: an inserted attribute violates an arithmetic constraint%) {
\end{lstlisting}  \vspace{-4mm} \begin{lstlisting}[language=Alloy,columns=flexible,escapechar=\%]
%\quad\quad\quad\quad\quad%fact {!(%\synt{int-expr}$_{\textsf{co}^{pos}}$ \textsf{co}$^{right}$%)}} }
\end{lstlisting}\vspace{-4mm}  \\ \hline
\synt{db-write} ::= 

\lit*{UPDATE} \synt{id}

\lit*{SET} \synt{id}$_{att}$\lit*{=}\synt{int-expr} 

\lit*{WHERE} \synt{db-cond} \lit*{;}
& \vspace{-4mm} \begin{lstlisting}[language=java,columns=flexible,escapechar=\%]
if (%no exception thrown in path for this UPDATE%) {
\end{lstlisting}  \vspace{-4mm} \begin{lstlisting}[language=Alloy,columns=flexible,escapechar=\%]
%\quad\quad%sig %\textsf{freshAlloyVar}% in %\synt{id}% {}
%\quad\quad%fact { all e:%\textsf{alloyName}% | (alloyOf(%\synt{db-cond}%,%\synt{id}%,e) && (one y:%\textsf{freshAlloyVar}% | 
%\quad\quad\quad\quad% y.%\textsf{att}%%$_{*-\{\synt{id}_{att}\}}$%= e.%\textsf{att}%%$_{*-\{\synt{id}_{att}\}}$% && y.%\synt{id}$_{att}$% = alloyOf(%\synt{int-expr}%,%\synt{id}%,e)))
%\quad\quad% || (!(alloyOf(%\synt{db-cond}%,%\synt{id}%,e)) && (one y:%\textsf{freshAlloyVar}% | equal%\synt{id}%[y,e])) }
%\quad\quad%fact { all y:%\textsf{freshAlloyVar}% | one e:%\textsf{alloyName}% | (alloyOf(%\synt{db-cond}%,%\synt{id}%,e) &&
%\quad\quad\quad\quad% y.%\textsf{att}%%$_{*-\{\synt{id}_{att}\}}$%= e.%\textsf{att}%%$_{*-\{\synt{id}_{att}\}}$% && y.%\synt{id}$_{att}$% = alloyOf(%\synt{int-expr}%,%\synt{id}%,e)))
%\quad\quad% || (!(alloyOf(%\synt{db-cond}%,%\synt{id}%,e)) && (one y:%\textsf{freshAlloyVar}% | equal%\synt{id}%[y,e])) }
%\quad\quad%fact { all disj a,b:%\textsf{freshAlloyVar}% | !(a.%\textsf{pk}% = b.%\textsf{pk}%) }  // Primary key is not violated
%\quad\quad%// If %\synt{id}$_{att}$ = \textsf{fk}$_{i}$%, updated rows should still reference a row		
%\quad\quad%fact { all a:%\textsf{freshAlloyVar}% | one b:%\textsf{fk}$_i^{tab}$% | a.%\synt{id}$_{att}$% = b.%\textsf{fk}$_i^{pk}$% }
%\quad\quad%// If %\synt{id}$_{att}$% = %\textsf{pk}%, none of the updated rows should have been referenced by another row
%\quad\quad%fact { all e:%\textsf{alloyName}% | no f:%\textsf{ifk}%%$_*^{tab}$% | (alloyOf(%\synt{db-cond}%,%\synt{id}%,e) && e.%\synt{id}$_{att}$% = f.%\textsf{ifk}%%$_*^{att}$% ) }			
\end{lstlisting}\vspace{-4mm} \begin{lstlisting}[language=java,columns=flexible,escapechar=\%]
} else { if (%Exception thrown in path: update on primary key leads to duplicate primary keys%) {
\end{lstlisting}  \vspace{-4mm} \begin{lstlisting}[language=Alloy,columns=flexible,escapechar=\%]
%\quad\quad\quad\quad\quad%fact { some disj a,b:%\textsf{alloyName}% | ((alloyOf(%\synt{db-cond}%,%\synt{id}$_{tab}$%,a)
%\quad\quad\quad\quad\quad%    && alloyOf(%\synt{db-cond}%,%\synt{id}$_{tab}$%,b) && (alloyOf(%\synt{int-expr}%,%\synt{id}%,a) 
%\quad\quad\quad\quad\quad%   = alloyOf(%\synt{int-expr}%,%\synt{id}%,b))) || (!(alloyOf(%\synt{db-cond}%,%\synt{id}$_{tab}$%,a)) 
%\quad\quad\quad\quad\quad%   && alloyOf(%\synt{db-cond}%,%\synt{id}$_{tab}$%,b) && (a.%\synt{id}$_{att}$% = alloyOf(%\synt{int-expr}%,%\synt{id}%,b))) ) }
\end{lstlisting}\vspace{-4mm} \begin{lstlisting}[language=java,columns=flexible,escapechar=\%]
%\quad%%\quad%%\quad%} else if (%Exception thrown in path: trying to update the primary key of a referenced row%) {
\end{lstlisting}  \vspace{-4mm} \begin{lstlisting}[language=Alloy,columns=flexible,escapechar=\%]
%\quad\quad\quad\quad\quad%fact { some a:%\textsf{alloyName}% | alloyOf(%\synt{db-cond}%,%\synt{id}$_{tab}$%,a)
%\quad\quad\quad\quad\quad\quad%&& (some %\textsf{ifk}$_j^{tab}$% | (b.%\textsf{ifk}$_j^{att}$% =a.%\synt{id}$_{att}$%)) }
\end{lstlisting}\vspace{-4mm} \begin{lstlisting}[language=java,columns=flexible,escapechar=\%]
%\quad%%\quad%%\quad%} else if (%Exception thrown in path: update on foreign key let row without row to reference%) {
\end{lstlisting}  \vspace{-4mm} \begin{lstlisting}[language=Alloy,columns=flexible,escapechar=\%]
%\quad\quad\quad\quad\quad%fact { some a:%\textsf{alloyName}% | alloyOf(%\synt{db-cond}%,%\synt{id}$_{tab}$%,a) 
%\quad\quad\quad\quad\quad\quad%&& (no b:%\textsf{fk}$_i^{tab}$% | b.%\textsf{fk}$_i^{pk}$%=alloyOf(%\synt{int-expr}%,%\synt{id}%,a)) }
\end{lstlisting}\vspace{-4mm} \begin{lstlisting}[language=java,columns=flexible,escapechar=\%]  
%\quad%%\quad%%\quad%} else if (%Exception thrown in path: an inserted attribute violates an arithmetic constraint%) {
\end{lstlisting}  \vspace{-4mm} \begin{lstlisting}[language=Alloy,columns=flexible,escapechar=\%]
%\quad\quad\quad\quad\quad%fact { some a:%\textsf{alloyName}% | alloyOf(%\synt{db-cond}%,%\synt{id}$_{tab}$%,a) 
%\quad\quad\quad\quad\quad\quad%&& !(alloyOf(%\synt{int-expr}%,%\synt{id}%,a))%\textsf{co}$_i^{right}$%)}} }
\end{lstlisting} \vspace{-6mm} \\ \hline
\synt{db-write} ::= 

\lit*{DELETE FROM} \synt{id}

\lit*{WHERE} \synt{db-cond} \lit*{;}

& \vspace{-4mm} \begin{lstlisting}[language=java,columns=flexible,escapechar=\%]
if (%no exception thrown in path for this DELETE%) { 
\end{lstlisting}  \vspace{-4mm} \begin{lstlisting}[language=Alloy,columns=flexible,escapechar=\%]
%\quad\quad%sig %\textsf{freshAlloyVar}% in %\synt{id}% {}
%\quad\quad%fact {%\textsf{freshAlloyVar}% = %\textsf{alloyName}% - {e:%\textsf{alloyName}% | alloyOf(%\synt{db-cond}%,%\synt{id}$_{tab}$%,e)}} 
%\quad\quad%// Not trying to delete a referenced row
%\quad\quad%fact {all e:%\textsf{alloyName}% | no f:%\textsf{ifk}$_j^{tab}$% | (alloyOf(%\synt{db-cond}%,%\synt{id}$_{tab}$%,e) && e.%\textsf{pk}% = f.%\textsf{ifk}$_j^{att}$%)}			
\end{lstlisting}\vspace{-4mm} \begin{lstlisting}[language=java,columns=flexible,escapechar=\%]
} else { // Trying to delete a referenced row 
\end{lstlisting}  \vspace{-4mm} \begin{lstlisting}[language=Alloy,columns=flexible,escapechar=\%]
%\quad\quad%fact {some e:%\textsf{alloyName}% | alloyOf(%\synt{db-cond}%,%\synt{id}$_{tab}$%,e) && (some f:%\textsf{ifk}$_j^{tab}$% | e.pk = f.%\textsf{ifk}$_j^{att}$%) } } }
\end{lstlisting} \vspace{-6mm} \\ \hline
\end{tabular}
\label{dbwrite}
\end{table*}

\lstset{language=Alloy,frame=none}
\begin{table*}
\centering
\caption{Translation of SimpleDB expressions and conditions into Alloy}
\begin{tabular}{|p{0.30\textwidth}|p{0.65\textwidth}|} \hline
\textbf{Parameters}& \textbf{alloyOf(Paramaters)} \\ \hline
\synt{id} & \textsf{alloyName} \\ \hline
$n \in \mathbb{N}$ & n \\ \hline
\lit*{(}\synt{int-expr}$_1$ \{\lit*{+}|\lit*{-}|\lit*{*}|\lit*{/}\} \synt{int-expr}$_2$\lit*{)} & (alloyOf(\synt{int-expr}$_1$)).\{add | sub | mul | div\}[alloyOf(\synt{int-expr}$_2$)]  \\ \hline
\lit*{(} \lit*{-} \synt{int-expr} \lit*{)}  & (- (alloyOf(\synt{int-expr}))) \\ \hline
\synt{id}\lit*{.HEAD} & \textsf{alloyName}.head \\ \hline
\synt{id}$_{tab}$\lit*{(}\synt{id}$_{att}$\lit*{)} &\{a:\textsf{alloyName}$_{tab}$ | all b:\textsf{alloyName}$_{tab}$ | a.\textsf{pk}$_{tab}$<=b.\textsf{pk}$_{tab}$\}.\synt{id}$_{att}$ \\ \hline
\lit*{NIL} & Nil \\ \hline
\synt{id}\lit*{.TAIL} & \textsf{alloyName}.tail \\ \hline
\lit*{TRUE} & (0=0) \\ \hline
\lit*{FALSE} & (0=1) \\ \hline
\lit*{(}\synt{cond}$_1$ \{\lit*{\&\&}|\lit*{||}\} \synt{cond}$_2$\lit*{)}  &  (alloyOf(\synt{cond}$_1$) \{\lit*{\&\&}|\lit*{||}\} alloyOf(\synt{cond}$_2$))  \\ \hline
\lit*{(} \lit*{!} \synt{cond} \lit*{)} & (!(alloyOf(\synt{cond}))  \\ \hline
\lit*{(}\synt{int-expr}$_1$ \{\lit*{<}|\lit*{=}|\lit*{>}\} \synt{int-expr}$_2$\lit*{)} & ((alloyOf(\synt{int-expr}$_1$)) \{\lit*{<}|\lit*{=}|\lit*{>}\} (alloyOf(\synt{int-expr}$_2$))) \\ \hline
\lit*{(}\synt{id}\lit*{=NIL}\lit*{)} & (\textsf{alloyName}=Nil) \\ \hline
 \multicolumn{2}{|c|}{\textbf{alloyOf(x,y,z) behaves in a similar way to alloyOf(x) except in the two following cases}}  \\ \hline
\synt{id}, table, row & \textbf{if} (table contains \synt{id}) \textbf{then} row.\synt{id} \textbf{else} \textsf{alloyName}\\ \hline
\lit*{(}\synt{id}\{\lit*{<}|\lit*{=}|\lit*{>}\} \synt{int-expr}\lit*{)}, table, row & (row.\synt{id} \{\lit*{<}|\lit*{=}|\lit*{>}\} (alloyOf(\synt{int-expr},table,row))) \\ \hline
\end{tabular}
\label{alloyOf}
\end{table*}

\begin{table*}
\centering
\caption{Abbreviations list and details}
\begin{tabular}{|p{0.15\textwidth}|p{0.80\textwidth}|} \hline
\textbf{Abbreviation}& \textbf{Meaning} \\ \hline
\textsf{freshAlloyVar} & A new Alloy variable name that has still not been used in the Alloy code generated so far. \\ \hline
\textsf{alloyName}$_{subscript}^{superscript}$ & \textbf{if} (\synt{id}$_{subscript}^{superscript}$ refers to a database table name) \textbf{then} The symbolic variable that represents the current content of table \synt{id}$_{subscript}^{superscript}$) \textbf{else} The symbolic variable that represents the current content of the SimpleDB variable \synt{id}$_{subscript}^{superscript}$  \\ \hline
\textsf{att}$_i$ & Name of the i$_{th}$ attribute in the list of attributes of table \synt{id} \\ \hline
\textsf{pk}$_{subscript}^{superscript}$ & Name of the primary key attribute of table \synt{id}$_{subscript}^{superscript}$. \\ \hline
\textsf{pk}$^{pos}$ & Position of primary key in the list of attributes of table \synt{id} \\ \hline
\textsf{fk}$_i^{tab}$ & Name of the table referenced by the i$_{th}$ foreign key in the list of foreign keys of table \synt{id} \\ \hline
\textsf{fk}$_i^{pk}$ & Name of the primary key attribute of the table referenced by the i$_{th}$ foreign key in the list of foreign keys of table \synt{id} \\ \hline
\textsf{fk}$_i^{pos}$ & Position of the foreign key attribute, declared by the i$_{th}$ foreign key in the list of table \synt{id}, in the list of attributes of table \synt{id} \\ \hline
\textsf{fk}$_i$ & Name of the foreign key attribute declared by the i$_{th}$ foreign key in the list of table \synt{id} \\ \hline
\textsf{ifk}$_i^{tab}$ & Name of the table where is declared the i$_{th}$ foreign key referencing table \synt{id} in the whole schema \\ \hline
\textsf{ifk}$_i^{att}$ & Name of the foreign key attribute declared by the i$_{th}$ foreign key referencing table \synt{id} in the schema\\ \hline
\textsf{co}$_i^{pos}$ & Position of the attribute constrained by the i$_{th}$ arithmetic constraint declared in table \synt{id}  \\ \hline
\textsf{co}$_i^{right}$ & Right part of the i$_{th}$ arithmetic constraint declared in table \synt{id} (i.e. right part of "a>0" is ">0") \\ \hline
\multicolumn{2}{|c|}{\textsf{xx}$_{*}$ means "for each \textsf{xx}$_i$" and \textsf{xx}$_{*-\{y\}}$ means "for each \textsf{xx}$_i$ except from y"}   \\ \hline
\end{tabular}
\label{abbrev}
\end{table*}

\section{Experimental evaluation}
\label{experimentation}

\begin{table*}
\centering
\caption{Experimental evaluation statistics}
\begin{tabular}{|c|l|l|p{1cm}p{1cm}|p{1.4cm}p{1.4cm}|p{1.6cm}p{1.6cm}|} \hline
\multirow{2}{*}{\textbf{Model}}  & \multirow{2}{*}{\textbf{Code lines}} & \multirow{2}{*}{\textbf{Tested paths}} & \multicolumn{2}{|c|}{\textbf{Symbolic variables}} & \multicolumn{2}{|c|}{\textbf{Relational constraints}}  & \multicolumn{2}{|c|}{\textbf{Constraints solving time}} \\ \cline{4-9}
& & & \textbf{Min} & \textbf{Max} & \textbf{Min} & \textbf{Max} & \textbf{Min} & \textbf{Max} \\ \hline
1 & 85 & 8 & 18 & 103 & 30 & 140 & 132 ms & 4,620 ms \\ \hline
2 & 65 & 8 & 4 & 47 & 17 & 131 & 309 ms & 262,320 ms \\ \hline
3 & 71 & 3 & 21 & 47 & 35 & 76 & 118 ms & 2,215 ms \\ \hline
\end{tabular}
\label{stat}
\end{table*}

The algorithm proposed in this work has been prototyped and evaluated experimentally using three SimpleDB test models. For each program, several execution paths have been symbolically executed by our algorithm. For each path, we asked the Alloy analyzer \cite{alloy-book} to find several solutions for the constraints generated by the algorithm. The evaluation process was completed by checking that each of these computed solutions was actually input data with respect to which the experimented program was guaranteed to follow the selected path. Both the SimpleDB test models and the tested paths were carefully selected to offer a good coverage of the promised abilities of our algorithm.  The Alloy analyzer is a program which allows to solve Alloy constraints in order to find structures that satisfy them. Basically, it transforms the set of relational constraints into an equivalent set of boolean constraints, and solves them using a SAT solver. The main statistics measured during the experimentation process for each of the three models are synthesized in table \ref{stat}. The whole material used and produced during the experimentation process, including the source code of the SimpleDB test models and the Alloy constraints generated by our algorithm, as well as all the performance-related information gathered during tests can be found on the web\footnote{See \url{http://info.fundp.ac.be/~mmr/issta13}}.

The first SimpleDB test model contains eighty-five lines of SimpleDB code that performs repeated manipulations of integers and lists using assignment, If and While statements. First, two lists are loaded and their size are compared. The program also reads as much integers from the outside world as the number of elements in the shortest of the two lists. A third list is created using these integers in the inverted order of the one in which they were read. If the two first lists have the same size, the elements of the three lists are compared. If the second list is an inverted version of the first one and if the third list is a copy of the first list where the value of each element was doubled, then the three lists are inserted into a database table. The database schema and the way the lists are inserted in this database constrain the elements in the first list to be different from each other and their value to range between one and five. The eight paths selected for this first test model were chosen so that only input lists with particular size and content allow the path to be followed during program execution. Between eighteen and one hundred and three symbolic variables and between thirty and one hundred and forty relational constraints were generated by the algorithm for each of the tested paths.

The second SimpleDB test model contains seventy five lines of SimpleDB code that performs repeated reads and writes in a database containing four tables that represent customers making purchases of products. Customers with few purchases are prospect customers. First, the program inserts a new customer and new purchases into the database. Then it computes the total number and cost of the purchases made by each customer, as well as the total number of purchases for each product, and then updates the corresponding customer and product attributes. All unpurchased products are deleted, and the name of the customers having made no purchase is changed. Customers whose total count and cost of purchase is lower than two are registered as prospect customers. Finally, a product is replaced by another one in every purchase details, and this product is deleted. The eight paths experimented over this program were carefully selected to test the generation of correct constraints for most possible behaviors for SQL statements over different kinds of tables structures and constraints. Between four and forty seven symbolic variables and between seventeen and one hundred and thirty one relational constraints were generated by the algorithm for each of the tested paths.

The third SimpleDB test model contains seventy one lines of SimpleDB code that mixes SQL statements with imperative code and uses SQL transactions. The database contains two tables that represent authors  writing theater plays. The code contains two transactions. During a first transaction, some authors are added and some removed from the database. During a second transaction, plays are added for the previously added authors, and some statistics are computed for each author. If a database schema constraint is violated by a SQL statement in one of the two transactions, this whole transaction is cancelled and the database is roll- backed to state it was when the transaction was launched. The selected paths for this third test model focus on the transaction management and contain a path in which both transactions are commited, one in which the first transaction is committed but the second is rolled back due to a foreign-key constraint violation, and a third one in which both transactions are partially executed but rolled back, also due to a foreign-key constraint violation in each of the transactions. Between twenty one and forty seven symbolic variables and between thirty five and seventy six relational constraints were generated for each of the tested paths.

For each of the tested paths among the three examples, every solution of the constraints generated by our algorithm provided a correct set of program inputs leading to the execution of the path. Concerning the infeasible path of the second test model, the Alloy analyzer was able to detect that the generated relational constraints were not satisfiable and did not proceed with SAT solving. Performance of the constraint solving process was acceptable in most cases, except for the longest paths involving many SQL statements, where it took up to four minutes to solve the constraints on a recent dual core x86 processor with 8GB of memory. 

\section{Conclusion and related work}
\label{conclusion}

In this work, we have proposed and detailed a complete algorithm to execute symbolically database programs. Given a data\-base program and an execution path in this program, the algorithm uses static analysis to generate a symbolic variable for each potential value taken by a program variable or database table before and during the path execution. It generates as well an Alloy relational constraints model constraining these symbolic variables to guarantee the execution of the considered path. Any solution to these produced constraints describes input data for the program, including an initial content for the database, with respect to which the program can be executed and is guaranteed to follow the considered execution path. Given a set of execution paths to test in the database program, satisfying a given code coverage criterion, the proposed algorithm can be used to generate inputs for each path in the set. These inputs can then be used in turn as adequate test data for structural white-box testing of the program. 

An early approach that has considered test data generation for imperative programs interacting with a relational SQL database is \cite{3}. The paper proposes to transform the program, thereby inserting new variables representing the data\-base structure, and translating all SQL statements and integrity checks into imperative program code. Classical white-box testing approaches can then be applied to the modified program. In \cite{13}, the authors propose an algorithm for testing an imperative program performing SELECT queries on a relational SQL database, based on a simultaneous concrete and symbolic (concolic) dynamic execution of some of its execution paths. Concolic execution runs the program on random input data and on a randomly populated input database. Given the dynamic exploration of an execution path of the program, the authors model and solve the problem of finding other inputs, allowing to explore dynamically another execution path, as a set of integer and string constraints over the quantity and field contents of the records in the database and over the input variables of the tested program. These constraints must be combined with the constraints derived from the database schema. In \cite{14}, authors adopt a similar concolic approach where the program is executed on a parameterized mock database. In \cite{15} and \cite{16}, authors adapt this approach to testing of programs running on an existing database, so that input test data can be selected in this database instead of being generated from scratch. In \cite{17}, the same concolic approach is applied considering advanced code coverage criteria. Compared to all of these approaches, our approach does not need to transform the original program, offers a clean modeling of the problem as a single relational constraints system generation problem, and allows to account for Insert, Update, and Delete statements, as well as transactions management primitives, that are commonly used in database applications. On the other hand, our approach only considers static SQL where the concolic ones allow to account for dynamic SQL. Finally, translation between database schemas/programs and Alloy models has already been considered in other contexts \cite{bdToAlloy,programToAlloy}. 

In future work, we intend to make our technique able to generate inputs for more complex interaction scenarios between databases and programs. First, it would be relevant to evaluate how and up to which extent the symbolic execution mechanism proposed here for simple SQL statements and simple relational database schemas can be generalized to more elaborate ones. Secondly, it should be investigated how dynamic SQL can be integrated with our approach, possibly relying on static analysis \cite{23,static-injection} or on concolic execution. Thirdly, it happens frequently that SQL statements have a non-deterministic behavior, either because the underlying DBMS executing the statement behaves non-deterministically, or because the database is modified concurrently by several programs. Whether and how the approach proposed here can encompass such non-deterministic behaviors remains a topic for further research. 

Finally, our approach allows to be used with respect to any classical code coverage criterion based on the notion of an execution path. Nevertheless, several works \cite{18,22,19,20,21} propose test adequacy criteria particularly adapted to the testing of database-driven programs. Integrating such particular coverage criteria into our constraint-based approach is a topic of ongoing research. \\

This work is unpublished work preliminary to \cite{moi-cstva,moi-scam13,moi-tech-report-scp}.

\section{Acknowledgments}

This work has been funded by the Belgian Fund for Scientific Research (F.R.S.-FNRS). The authors would like to thank Vincent Englebert for useful discussions.

\bibliographystyle{abbrv}
\bibliography{issta13} 

\end{document}